\documentclass[11pt]{article}
\usepackage{graphicx}
\topmargin - 2cm
\oddsidemargin - 0.1cm
\evensidemargin - 5cm

\newtheorem{thm}{Theorem}[section]

\newtheorem{lem}[thm]{Lemma}
\newtheorem{cor}[thm]{Corollary}

\newtheorem{remark}[thm]{Remark}

\begin{document}

\title{\small{{\bf MULTIVARIATE GENERALIZATIONS OF THE $q$--CENTRAL LIMIT THEOREM }}}
\author{Sabir Umarov$^{1}$, Constantino Tsallis$^{2}$ }  
\date{}
\maketitle
\begin{center}
$^{1}$ {\it Department of Mathematics and Statistics\\ 
University of New Mexico, 
Albuquerque, NM 87131, USA}\\

$^2$ {\it Centro Brasileiro de Pesquisas Fisicas \\ 
Xavier Sigaud 150, 22290-180 Rio de Janeiro-RJ, Brazil}
\end{center}

\begin{abstract} 
We study multivariate generalizations of the $q$-central limit theorem, a generalization of the classical central limit theorem consistent with nonextensive statistical mechanics.  
Two types of generalizations are addressed, more precisely the {\it direct} and {\it sequential} $q$-central limit theorems are proved. 
Their relevance to the asymptotic scale invariance of some specially correlated systems is studied.
A $q$-analog of the classic weak convergence is introduced and its equivalence to the $q$-convergence 
is proved for $q>1$.

\end{abstract}


\section{INTRODUCTION}
We studied $q$-analogs of the classical central limit theorem and 
$\alpha$-stable distributions for $q$-independent random variables 
in \cite{UmarovTsallisSteinberg,UmarovTsallisGellmannSteinberg1,UmarovTsallisGellmannSteinberg2}. The results that we obtained there are 
consistent with nonextensive statistical mechanics \cite{Tsallis1988,GellMannTsallis2004,BoonTsallis2005} and represent limit states
of sequences of globally correlated systems. These limit states are described by $q_{-1}$-Gaussian distributions,
where 
\[
q_{-1}=\frac{3q-1}{q+1}\,,
\]
where the index $q$ characterizes the correlation. 
The particular case $q=1$, for which  $q_{-1}=1$, corresponds to independent random variables, and exibits the classic central limit theorem. If $q \neq 1$ then $q_{-1} \neq q$
and there exist series of $q$-CLTs depending on the type of correlation (see \cite{UmarovTsallisSteinberg}). The main tool used for the study of 
the $q$-central limit theorems is the $q$-Fourier transform introduced as 
\[
F_q[f]=\int_R f(x) \otimes_q e_q^{ix \xi} dx = \int_R f (x) e_q^{ix \xi [f(x)]^{q-1}  } dx,
\]
where $f:R \rightarrow R^1_+$  
is a density function, and $e_q(x)$ is the $q$-exponential (for elements of the $q$-algebra used in the
current paper, see \cite{qborges1,qborges,qnivanen}). $F_q$ is a nonlinear transfomation if $q \neq 1$.

In the present paper we will study a multivariate analog of the $q$-CLT. We will introduce two generalizations, denoted  {\it direct} and {\it sequential}, of the $q$-Fourier transform to the
multidimensional case. In turn, they lead to the two generalizations of the $q$-CLT that we address here.
Finally, we apply the multivariate $q$-CLT to the study of the scale invariance of statistical quantities with a finite $q$-mean 
and arbitrary $(2q-1)$-variance matrix.

\section{PRELIMINARIES}

Let  $(\Omega, \mathcal{F}, P)$ be a probability space and 
$X:\Omega \rightarrow R^d$ be a $d$-dimensional random vector with a joint density function $f_X(x)=f(x), \, x=(x_1,...,x_d) \in R^d$. 
We introduce the associated {\it escort density} \cite{beck}
$$f_q(x)=\frac{[f(x)]^q}{\nu_q(f)},$$ where $\nu_q(f)= \int_{R^d} [f(x)]^q  \, dx$. 
By definition, the $q$-mean of $X$ is a vector $\mu_q(X) = \mu_q =(\mu_{q,1},...,\mu_{q,d})^T$, where $\mu_{q,j} = \mu_{q,j}(X_j) = 
\int_{R^d} x_j f_q(x)\,dx, \, \, j=1,..,d$, and the $q$-covariance matrix 
$\Sigma_q$ is a matrix with entries $\sigma_{q,i,j}=\int_{R^d} (x_j-\mu_{q,j})(x_k - \mu_{q,k}) f_q(x)\,dx$. 
By the direct $d$-dimensional $q$-Fourier transform we understand the integral
\begin{equation}
\label{d-Fourier}
\mathcal{F}_q[f](\xi)=\int_{R^d} f(x) \otimes_q e_q^{ix \xi} dx = \int_{R^d} f (x) e_q^{ix \xi [f(x)]^{q-1}  } dx, \, \, 
\end{equation}
where $\xi \in R^d$ and $ix \xi [f(x)]^{q-1} = i [f(x)]^{q-1} \sum_{k=1}^d x_k \xi_k $, subject to the existence of the integral. 
Let $q=1$. Then, (\ref{d-Fourier}) identifies the 
classic Fourier transform
\[
\mathcal{F}[f](\xi)=\int_{R^d} f(x) e^{ix \xi} dx. 
\]
Moreover, in this case $\mathcal{F} = F_1 \circ ... \circ F_d$, where $F_j, \, j=1,...,d,$ 
are the one-dimensional Fourier transforms with respect to the variables $x_j, j=1,...,d,$ and the symbol
"$\circ$" means the composition of operators $F_j, \, j=1,...,n.$
This property fails if $q \neq 1$. That is, generically
\[
\mathcal{F}_q \neq F_{1,q} \circ ... \circ F_{d,q},
\]
where $F_{j,q}$ is the $q$-Fourier transform with respect to $x_j, \, \, j=1,...,d$.
Assume 
\begin{equation}
\label{qn}
q_n = \frac{2q + n(1-q)}{2 + n(1-q)}, \, \, n=0, \pm1, \pm 2,... 
\end{equation}
i.e., equivalently,
\begin{equation}
\label{qnn}
\frac{1}{1-q_n} = \frac{1}{1-q}+\frac{n}{2}, \, \, n=0, \pm1, \pm 2,... 
\end{equation}
Let $Q_k$ be an ordered set of numbers $Q_k=\{q_{k},q_{k+1},...,q_{k+d-1}\}$. The composition
\begin{equation}
\mathcal{F}_{Q_k}[f] (\xi)=F_{1, q_{k+d-1}}\circ ... \circ F_{d, q_k} [f(x_1,...,x_d)](\xi_1,...,\xi_d)
\end{equation}
is called a {\it $Q_k$-sequential Fourier transform} of a given function $f(x)$.
The direct and sequential Fourier transforms coincide, only if $q=1.$ In this case $\mathcal{Q}=\{1,...,1\}.$

A $d$-dimensional $q$-Gaussian with a covariance matrix $\Sigma$ is a function
\begin{equation}
\label{qgau}
G_q (\Sigma; x)=C_{d,q}(\beta)\, e_{q}^{- \beta (\Sigma x, x)},
\end{equation}
where $\beta $ is a positive real number, $\Sigma$ is a positively defined $d \times d$-matrix with 
entries $\sigma_{q^{'}, i,j}, \, \, i,j=1,...,d,$ and
$$C_{d,q}(\beta) = C_{d,q}(\beta, \Sigma) = (\int_{R^d} e_{q}^{-\beta (\Sigma x,x) } dx  )^{-1}$$ 
is the normalising constant. We denote $C_{d,q}=C_{d,q}(1, \Sigma).$ Clearly, $C_{d,q}(\beta)=C_{d,q}(1, \beta \Sigma).$ 
In the particular case of a diagonal covariance matrix $\Sigma$ with diagonal entries
$\sigma_{j,j}>0$, the $q$-Gaussian takes the form
\[
G_{q}(\Sigma; x)=C_{d,q}(\beta) \, e_{q}^{-\beta (\sigma_{1,1} |x_1|^2 +...+\sigma_{d,d} |x_d|^2 )}.
\]

Now we compute the value of $C_{d,q}(\beta)$ as a function of $q, \, \beta$ and $\Sigma$. First, assume $\beta=1$ and
$det(\Sigma)=|\Sigma|>0.$ 
There is an orthogonal matrix $U$ such that $\Sigma=U^{T}DU$ where $D$ is a diagonal matrix. Substituting $y=Ux$ in the 
integral
$
(C_{d,q})^{-1} = \int_{R^d} e_q^{-(\Sigma x, x)} dx
$
we have
\[
(C_{d,q})^{-1} = \int_{R^d} e_q^{-(D y, y)} dy = \frac{1}{\sqrt{|\Sigma|}} \int_{R^d} e_q^{-|z|^2} dz =
\frac{\omega_d I_{q,d}}{\sqrt{|\Sigma|}}, 
\]
where $\omega_d = \frac{2 \pi^{d \over 2}}{\Gamma ({d \over 2})}$ is the area of the surface of the 
unit sphere in $R^d$ and 
\begin{equation}
\label{iqd}
I_{q,d} = \int_0^{\infty} r^{d-1} e_q^{-r^2} dr.
\end{equation}
Note that $I_{q,d}<\infty$ if $q < 1 + {2 \over d}$. Hence,
\begin{equation}
\label{cq1}
C_{d,q} = \frac{\sqrt{|\Sigma|}}{\omega_d I_{q,d}}. 
\end{equation}

We also note that consequently integrating the
$d$-dimentional $q$-Gaussian, for $C_{d,q}$ with $\Sigma =I$ ($I$ is the identity matrix), we obtain the formula
\begin{equation}
\label{cq11}
C_{d,q} =  \frac{ (\frac{3-q_1}{2})^{d-1 \over 2}(\frac{3-q_2}{2})^{d-2 \over 2} ... (\frac{3-q_{d-1}}{2})^{1 \over 2}  }
{ C_{q}^{q-1} C_{q_1}^{q_1-1}...C_{q_{d-1}}^{q_{d-1}-1}  }, 
\end{equation}
with $q, q_1,..., q_{d-1}$ identified in (\ref{qn}), and $C_q$, the normalising constant of the one-dimensional Gaussian.

For an arbitrary $\beta > 0$, (\ref{cq1}) implies the following lemma.
\begin{lem}
For the normalising constant of a $q$-Gaussian $G_q (\Sigma; x)=C_{d,q} e_{q}^{- \beta (\Sigma x, x)}$, the formula
\begin{equation}
\label{cq}
C_{d,q}(\beta) = \frac{\beta^{d \over 2} \sqrt{|\Sigma|} \Gamma ({d \over 2})}{2 \pi^{d \over 2} I_{q,d}} = \beta^{\frac{d}{2}}C_{d,q}
\end{equation}
holds, where $\Gamma (\cdot) $ is the Euler's gamma-function and  $I_{q,d}$ is defined in (\ref{iqd}).
\end{lem}
\begin{lem}
\label{l1}
Let $q<1+{2 \over d}$ and $\Sigma$ be a symmetric positively defined $d \times d$ matrix with
a determinant $|\Sigma| > 0$. Then
\begin{equation}
\label{ftg}
\mathcal{F}_q [C_{d,q}(\beta) e_q^{- \beta (\Sigma x, x)}] (\xi) = 
\left( e_q^{-\frac{C_{d,q}^{2(q-1)}(\Sigma^{-1}\xi,\xi)}{4 \beta^{1+d-dq}}} \right)^{\frac{2-d(q-1)}{2}}.
\end{equation}
\end{lem}

{\it Proof.} First we assume that $\beta = 1$. Applying elementary properties of $q$-exponential and $q$-product we have
\begin{equation}
\label{1.1}
\mathcal{F}_q [ C_{d,q} e_q^{ -(\Sigma x, x) }] (\xi) = C_{d,q} \int_{R^d} e_q^{-(\Sigma x, x)+i x \xi C_{d,q}^{q-1}} dx.
\end{equation}
Further, by assumption, $\Sigma$ is a symmetric positive matrix. Therefore there exists an orthogonal matrix $U$ such that
$D=U^{-1} \Sigma U$ is a diagonal matrix. Assume $\delta_1,...,\delta_n$ are the diagonal entries of $D$. Then (\ref{1.1}) is reduced to
\begin{equation}
\label{1.2}
\mathcal{F}_q [ C_{d,q} e_q^{ -(\Sigma x, x) }] (\xi) = \frac{C_{d,q}}{\sqrt{|\Sigma|}} \int_{R^d} e_q^{|z|^2} 
\otimes_q e_q^{- {1 \over 4} C_{d,q}^{2(q-1} (D^{-1} U \xi, U \xi)} dz,
\end{equation}
where $|\Sigma| = \delta_1 ... \delta_n$ is the determinant of $\Sigma$. Again using the properties of $q$-exponential and $q$-product we compute the
right hand side of (\ref{1.2}):
\[
\mathcal{F}_q [ C_{d,q} e_q^{ -(\Sigma x, x) }] (\xi)= \frac{C_{d,q}}{\sqrt{|\Sigma|}} 
e_q^{- {1 \over 4} C_{d,q}^{2(q-1} (\Sigma^{-1} \xi, \xi)} \int_{R^d} e_q^{-|z|^2 (e_q^{- {1 \over 4} C_{d,q}^{2(q-1} (\Sigma^{-1} \xi, \xi)})^{q-1}} dz=
\]
\[
(e_q^{- {1 \over 4} C_{d,q}^{2(q-1} (\Sigma^{-1} \xi, \xi)})^{1- {d \over 2}(q-1)}.
\]
Finally, for $\beta \neq 1$, replacing $\Sigma$ by $\beta \Sigma$ and taking into account (\ref{cq}), we arrive at (\ref{ftg}).

\begin{cor}
\label{cormain}
Under the conditions of Lemma \ref{l1} 
\begin{equation}
\label{image}
\mathcal{F}_q [C_{d,q}(\beta) e_q^{-\beta(\Sigma x, x)}] (\xi) = 
            e_{q_1}^{-\alpha (\Sigma^{-1}\xi,\xi) },
\end{equation}
where 
\begin{equation}
\label{qd1}
q_1 = \frac{2q-d(q-1)}{2-d(q-1)}=1+\frac{2(q-1)}{2-d(q-1)}
\end{equation} 
and 
\begin{equation}
\label{alpha}
\alpha = \frac{(2-d(q-1))[C_{d,q}(\beta)]^{2(q-1)}  }{8 \beta}. 
\end{equation}
\end{cor}

Replacing $C_{d,q}(\beta)$ by its value in equation (\ref{cq}) we have
\[
\alpha = \frac{2-d(q-1)}{2^{2q+1} \beta^{1+d-dq} \pi^{d(q-1)}} \left( \frac{\sqrt{|\Sigma|}\Gamma({d \over 2})}{I_{q,d}} \right)^{2(q-1).} .
\]

Introduce the function 
$$
z^d(s)=\frac{2s-d(s-1)}{2-d(s-1)}=1+\frac{2(s-1)}{2-d(s-1)},
$$
and denote its inverse $z^{d,-1}(t)$, which reads
\[
z^{d,-1}(t)= \frac{2t+d(t-1)}{2+d(t-1)}=1+\frac{2(t-1)}{2+d(t-1)}.
\] 
One can easily verify that $z^d \bigl(\frac{1}{z^d(s)} \bigr)= {1 \over s}$ and $z^d ({1 \over s})={1 \over z^{d,-1}(s)}\,.$ 
Moreover, it is readely verified that $z(s)$ possesses the property 
\begin{equation}
\label{qdual2} 
(z^d(s))^{-1} + d z^{d,-1} (s) = 1+d. 
\end{equation}

Further, let us introduce the sequence $q^d_n = z^d (q^d_{n-1}), n=1,2,...,$  with a given $q^d_0 = q, \, q<1+{2 \over d}.$ 
We can extend the sequence $q_n$ for negative integers $n=-1,-2,...$ as well, putting
$q_{-n} = z^{-1}(q^d_{1-n}), n = 1, 2,...\,.$ 
It is not hard to verify that 
\begin{equation}
\label{qdn}
q^d_n = \frac{2q + d n(1-q)}{2 + d n(1-q)} = 1 + \frac{2(q-1)}{2-dn(q-1)} , \, \, n=0, \pm1, \pm 2,... 
\end{equation}
i.e., equivalently,
\begin{equation}
\label{qnn}
\frac{1}{1-q_n^d} = \frac{1}{1-q}+\frac{dn}{2}, \, \, n=0, \pm1, \pm 2,... 
\end{equation}
Note $q^d_n \equiv 1$ for all $n=0, \pm 1, \pm 2,...,$ if $q=1$ and $lim_{ n \rightarrow \pm \infty} q_n^d=1$ 
for all $q \neq 1.$

\begin{remark} 
Essentially the same mathematical structure has already appeared in a quite different, 
though possibly related, context: see Footnote of page 15378 of \cite{TsallisGellmannSato}. In the
one-dimensional case, i.e for $d=1$, the relationship (\ref{qdn}) is obtained in \cite{UmarovTsallisSteinberg}. For $n=1$, in the
$d$-simensional case, the relationship
\begin{equation}
\label{qd1}
q^d_1 = \frac{2q - d (q-1)}{2 - d (q-1)} = 1 + \frac{2(q-1)}{2-d(q-1)}, 
\end{equation}
between components of elliptically invariant Gaussians was recorded in \cite{VignatPlastino}.
\end{remark}

\begin{lem} The members of the sequence $q_n^d$ satisfy the following duality relations
\begin{equation}
\label{dualityrelation}
d q^d_{n-1} + \frac{1}{q^d_{n+1}} = 1+d, \, n=0, \pm1, \pm2,....
\end{equation}
\end{lem}

{\it Proof.} By elementary calculations we have
\[
1+d-dq^d_{n-1}=1+\frac{2d(1-q)}{2+d(n-1)(1-q)}= \frac{1}{\frac{2q +d(n+1)(1-q)}{2+d(n+1)(1-q)}}=\frac{1}{q^d_{n+1}} \,.
\]

\begin{remark}
In the one-dimensional case ($d=1$) the duality relations (\ref{dualityrelation}) recover the well known relationships 
(see, e.g., \cite{UmarovTsallisSteinberg,Tsallis2005,MoyanoTsallisGellmann2006})
\[
q_{n-1}+\frac{1}{q_{n+1}}=2,\, \, n=0,\pm 1,....
\]
\end{remark}

It follows from Corollary \ref{cormain} and the definition of the sequence $q^d_n$ the following statement.
\begin{lem} The series of mappings 
\begin{equation}
\label{Fright}
... \stackrel{\mathcal{F}_{q^d_{-3}}}{\rightarrow}\mathcal{G}_{q^d_{-2}} \stackrel{\mathcal{F}_{q^d_{-2}}}{\rightarrow} \mathcal{G}_{q^d_{-1}} 
\stackrel{\mathcal{F}_{q^d_{-1}}}{\rightarrow} \mathcal{G}_{q} \stackrel{\mathcal{F}_q}{\rightarrow} \mathcal{G}_{q^d_{1}} 
\stackrel{\mathcal{F}_{q^d_1}}{\rightarrow} \mathcal{G}_{q^d_{2}}  \stackrel{\mathcal{F}_{q^d_2}}{\rightarrow} ...  
\end{equation}
\begin{equation}
\label{Fleft}
... \stackrel{\mathcal{F}^{-1}_{q^d_{-3}}}{\leftarrow} \mathcal{G}_{q^d_{-2}} \stackrel{\mathcal{F}^{-1}_{q^d_{-2}}}{\leftarrow} \mathcal{G}_{q^d_{-1}}  
\stackrel{\mathcal{F}^{-1}_{q^d_{-1}}}{\leftarrow} \mathcal{G}_{q}  \stackrel{\mathcal{F}^{-1}_q}{\leftarrow} \mathcal{G}_{q^d_{1}}  
\stackrel{\mathcal{F}^{-1}_{q^d_1}}{\leftarrow} \mathcal{G}_{q^d_{2}}  \stackrel{\mathcal{F}^{-1}_{q^d_2}}{\leftarrow} ...   
\end{equation}
hold.
\end{lem}

\begin{lem}
\label{l2}
\begin{equation}
F_q[C_0 e_q^{-\beta (x^2+A) }](\xi) = C_1 e_{q_1}^{-\beta_1 (A + \alpha_1 \xi^2)},
\end{equation}
with $C_1=\frac{C_0 C_q}{\sqrt{\beta}}$, $\beta_1=\frac{(3-q)\beta}{2}$ and $\alpha_1 = \frac{C_0^{q-1}}{4 \beta^2}$.

\end{lem}

\begin{cor}
\label{Qkcor}
Let $\beta_j > 0, \, j=1,...d,$ be given numbers. Then
\begin{equation}
\label{fqk}
\mathcal{F}_{Q_k}[ C_0 e_{q_k}^{\beta_1 x_1^2 +...+ \beta_d x_d^2} ] = C_d e_{q_{k+d}}^{- \sum_{j=1}^d 
\frac{\gamma_{d-j+1}C_{d-j}^{q_{k+d-1}-1}} {4 \alpha_{d-j}\beta_j} \xi_j^2   },
\end{equation}
where 
\begin{equation}
\label{alfagama}
\alpha_m = \prod_{j=1}^m \frac{3-q_{k+j-1}}{2}, \, \, \,  \gamma_m=\prod_{j=m+1}^d \frac{3-q_{k+j-1}}{2}
\end{equation} 
and 
\[
C_m = \frac{C_0}{\sqrt{\beta_{d}...\beta_{d-j+1}}} \prod_{j=1}^m \frac{C^{q_{k+j-1}-1}_{q_{k+j-1}}}{(\frac{3-q_{k+j-1}}{2})^{m-j \over 2}}\, \, m=1,...,d.
\]
\end{cor}

\begin{remark}
It follows from (\ref{cq11}) that, if we take $C_0 = \frac{\sqrt{\beta_1 ... \beta_d}}{C_{d,q_k}}$ (in which case 
$C_0 e_{q_k}^{\beta_1 x_1^2 +...+ \beta_d x_d^2}$ is a Gaussian density), then $C_d = 1$ in (\ref{fqk}).
\end{remark}

\section{THE $q$-CONVERGENCE}

In this section we introduce the notion of the {\it weak $q$-convergence} of a given sequence of random
variables, and compare it with the multivariate generalization of the {\it $q$-convergence}. 
A sequence of random vectors $X_N$ is said to be $q$-convergent to a random vector $X$ if
$\mathcal{F}_q[X_N](\xi) \rightarrow \mathcal{F}_q[X](\xi)$ for every $\xi \in K$, where 
$K$ is arbitrary compact in $R^d$. In the one-dimensional case, this definition was introduced in \cite{UmarovTsallisSteinberg}. 
In terms of associated densities the $q$-convergence is equivalent to the locally uniform (by $\xi$) convergence
$\mathcal{F}_q[f_N](\xi) \rightarrow \mathcal{F}_q[f](\xi)$, where 
$f_N$ and $f$ are densities of $X_N$ and $X$ respectively. We denote the $q$-convergence by the symbol
$\stackrel{q}{\rightarrow}$.

Recall that a sequence $X_N$ is called weakly convergent 
to a random vector $X$ if
the sequence of corresponding distribution functions $F_{X_N}(x)$ converges to $F_{X}(x)$ at every point of
continuity \cite{Billingsley}. The weak convergence is usually denoted by ${\Rightarrow}$. 
The equivalent definition of the weak convergence in terms of associated densities is the
following: for any bounded continuous function $g(x)$, the sequence of integrals
$\int_R f_N(x)g(x)dx$ converges to $\int_R f(x)g(x)dx$ \cite{Billingsley}. For $q \neq 1$ denote by $W_q$ the set of functions 
$\phi$ continuous on $R^d$ and satisfying the condition $|\phi(x)| \leq C(1+|x|)^{\frac{q}{1-q}}, \, x \in R^d$.
A sequence of random vectors $X_N$ is called weakly $q$-convergent to a random vector $X$ if  
$\int_{R^d} f_N(x) d m_q \rightarrow \int_{R^d} f(x)d m_q $ for arbitrary measure $m_q$ defined as 
$dm_q()= \phi_q (x) dx,$ where $\phi_q \in W_q$. We denote the weak $q$-convergence by the symbol
$\stackrel{q}{\Rightarrow}$. 
Notice that, in the 1-dimensional case ($d=1$), any measure $m_q, \, \phi_q \in W_q$ is finite (i.e. $m_q(R^1)<\infty$) for all $q > 1$.  
If $d \geq 2$ then for $1<q<1+1/(d-1)$ a measure $m_q$ is finite (i.e. $m_q(R^d)<\infty$). 
If $0<q<1$, then there exists a function $\phi_q \in W_q$ such that  
the corresponding measure $m_q$ diverges (i.e. $m_q(R^d)=\infty$).

\begin{lem}
\label{weak}
If $q>1$ then $X_N \Rightarrow X_0 $ yields $X_N \stackrel{q}{\Rightarrow} X_0$. If $q<1$ then
$X_N \stackrel{q}{\Rightarrow} X_0$ yields  $X_N \Rightarrow X_0$. 
\end{lem}

The proof of this lemma follows from the obvious fact that $W_q$ is a subset of the set of bounded
functions if $q>1$, and $W_q$ contains the set of bounded functions  as a subset if $q<1$.
\vspace{.3cm}

The next lemma uses the classic notion of {\it tightness} of a sequence of probability measures. Recall that
a sequence of probability measures $\mu_N$ is called tight if, for an arbitrary $\epsilon >0$, there 
is a compact $K_{\epsilon}$ and an integer $N^{\ast}_{\epsilon}$ such that
$\mu_N(R^d \setminus K_{\epsilon}) < \epsilon$ for all $N \geq N_{\epsilon}^{\ast}$.

\begin{lem}
\label{tight}
Let $1<q<1+1/(d-1)$ if $d \geq 2$ and $1<q<2$ if $d=1$. Assume a sequence of random vectors $X_N$, 
defined on a probability space with a probability measure $P$, is
$q$-convergent. Then the sequence of associated probability measures $\mu_N = P(X^{-1})$ is tight.
\end{lem}

{\it Proof.} It suffices to prove the statement in the one-dimensional case. The multivariate extension can be obtained
using the analogous arguments proceeded in the classic case (i.e. in the case $q=1$) using the Cramer-Wold device (see,
for instance, \cite{Billingsley, Durrett}). Thus, assume that $1<q<2$ and $X_N$ is a $q$-convergent
sequence of random variables with associated densities $f_N$ and associated probability measures $\mu_N$.
We have
\[
\frac{1}{R}\int_{-R}^{R}(1-F_q[f_N](\xi)) d \xi = \frac{1}{R}\int_{-R}^{R}(1-\int_{R} f_N e_q^{i x \xi f_N^{q-1}} dx ) d \xi = 
\]
\begin{equation}
\label{tight1}
\int_{R} \left( \frac{1}{R}\int_{-R}^{R}(1- e_q^{i x \xi f_N^{q-1}} ) d \xi \right) d \mu_N(x).
\end{equation}
It is not hard to verify that
\begin{equation}
\label{tight2}
\frac{1}{R}\int_{-R}^{R} e_q^{i x \xi t} d \xi =  \frac{2 sin_{\frac{1}{2-q}} (R x(2-q) t)}{Rx(2-q)t}.
\end{equation}
It follows from (\ref{tight1}) and (\ref{tight2}) that 
\begin{equation}
\label{tight3}
\frac{1}{R}\int_{-R}^{R}(1-F_q[f_N](\xi)) d \xi =
2\int_{-\infty}^{\infty} \left( 1 - \frac{sin_{ \frac{1}{2-q} } (x(2-q)R f_N^{q-1})}{Rx(2-q)f_N^{q-1}} \right) d\mu_N(x).
\end{equation}
Since $1<q<2$ by assumption,  $\frac{1}{2-q} > 1$ as well. It is known \cite{UmarovTsallisGellmannSteinberg1} that for any $q^{'}>1$ the 
properties $sin_{q^{'}}(x) \leq 1$ and $(sin_{q^{'}}(x))/x \rightarrow 1, \, x \rightarrow 0$ hold. It follows from these facts 
and $\{x: |x|f_N^{q-1}(x) >\frac{2}{R(2-q)} \} \subset \{x: |x| >\frac{2}{R(2-q)} \}$ that
\[
\frac{1}{R}\int_{-R}^{R}(1-F_q[f_N](\xi)) d \xi \geq 2 \int_{|x|f_N^{q-1} \geq \frac{2}{R(2-q)}} \left( 1-\frac{1}{R|x|(2-q)f_N^{q-1}} \right) d\mu_N (x)
\]
\[
\geq 2 \int_{|x| \geq \frac{2}{R(2-q)}} \left( 1-\frac{1}{R|x|(2-q)} \right) d\mu_N (x) \geq 2 \mu_N \left( |x| \geq \frac{2}{R(2-q)} \right).
\]
Further taking into account the $q$-convergence of $f_N$ to $f_0$ and taking $R$ small enough we obtain
\[
\mu_N \left( |x| \geq r \right) \leq  \frac{1}{2R}\int_{-R}^{R}(1-F_q[f_0](\xi)) d \xi < \epsilon,
\] 
where $r=\frac{2}{R(2-q)},$ and $\epsilon $ is an arbitrary positive number.

\vspace{.3cm}

Introduce for $t \in [0,1]$ the function  
\begin{equation}
\label{dt}
D_q(t)= D_q(t; a)=t e_q^{i a t^{q-1}}= t(1 + i (1-q) a t^{q-1})^{\frac{1}{1-q}},
\end{equation}
where $a$ is a fixed real number. Obviously $D_q(t)$ is continuous and differentiable
in the interval $(0,1)$. In accordance with the classical Lagrange average theorem for any 
$t_1, t_2, \, \, 0 \leq t_1 < t_2 \leq 1$ there exists a number
$t_{\ast}, \, \, t_1 < t_{\ast} < t_2$ such that 
\begin{equation}
\label{ev}
D_q(t_1)-D_q(t_2) = D_q^{'}(t_{\ast}) (t_1 - t_2),
\end{equation} 
where $D_q^{'}$ means the derivative of $D_q(t)$ with respect to $t$.

Consider the following Cauchy problems for the two Bernoulli equations
\begin{equation}
\label{bern1}
y^{'} - \frac{1}{t} y = \frac{ia(1-q)}{t} y^q, \, \, y(0)=0, \, \hspace{2cm} (q>1)
\end{equation}
and
\begin{equation}
\label{bern2}
y^{'} = \frac{1}{t^q} y^q, \, \, y(0)=(ia)^{\frac{1}{1-q}}. \, \hspace{2.3cm}  (q<1)
\end{equation}
It is not hard to verify that $D_q(t)$ is a solution to the problems (\ref{bern1}) and (\ref{bern2}) with $a= x \xi$
for $q>1$ and $q<1$ respectively.

\begin{lem}
For $D^{'}_q (t),$  $q \neq 1$, the estimate
\begin{equation}
\label{est}
|D_q^{'}(t;a)| \leq C (1 + |a|)^{-\frac{q}{q-1}}, \, \, t \in (0,1], \, a \in R^1,   
\end{equation}
holds, where $C$ does not depend on $t$. 
\end{lem}

{\it Proof.} Let $q<1.$ Then it follows from (\ref{bern2}) that

\[
|y^{'}(t)| =  t^{-q} |y^q| = |e_q^{i a t^{q-1}}|^q = (e_q^{(1-q)a^2t^{2(q-1)}})^{\frac{q}{2}} \leq
\]
\[
C(1+|a|)^{\frac{q}{1-q}}, \, \, t \in (0,1].
\]
Now assume $q>1.$ Then (\ref{dt}) and (\ref{bern1}) imply
\[
|y^{'}(t)| \leq t^{-1}|y + i a (1-q)y^{q}| = |e_q^{i a t^{q-1}} + i a (1-q)(e_q^{i a t^{q-1}})^q|=
\]
\[
|1+ia(1-q)t^{q-1}|^{-\frac{q}{q-1}}  \leq C (1+|a|)^{-\frac{q}{q-1}}, \, t \in (0,1].
\]

\begin{thm}
Let $1<q<1+\frac{1}{(d-1)}$ if $d \geq 2$ and $q>1$ if $d=1$. Let a sequence of random vectors $X_N$ be weakly $q$-convergent to a random vector 
$X$. Then $X_N$ is $q$-convergent to $X$.
\end{thm}

{\it Proof.} 
Assume $X_N$ with associated densities $f_N$ is weakly $q$-convergent to $X$ with an associated density $f$.
The difference $\mathcal{F}_q [f_N](\xi) - \mathcal{F}_q [f_N](\xi)$ can be written
\begin{equation}
\label{difference}
\mathcal{F}_q [f_N](\xi) - \mathcal{F}_q [f_N](\xi)
       =  \int_{R^d} \left( D_q(f_N(x)) - D_q( f(x) ) \right) dx, 
\end{equation}
where $D_q(t)=D_q(t;a)$ is identified in (\ref{dt}) with $a = x \xi$.  It follows from (\ref{ev}) and (\ref{est}) that
\[
|\mathcal{F}_q [f_N](\xi) - \mathcal{F}_q [f_N](\xi)| \leq C \int_{R^d} | (1+|x|)^{-\frac{q}{q-1}} \left(f_N(x)- f(x) \right)|dx,
\]
which yields $\mathcal{F}_q [f_N](\xi) \rightarrow \mathcal{F}_q [f_N](\xi)$ for all $\xi \in R^d$.

\begin{thm}
Let $1<q<1+\frac{1}{(d-1)}$ if $d \geq 2$ and $1<q<2$ if $d=1$. Assume a sequence of random 
vectors $X_N$ with the associated densities $f_N$ is $q$-convergent to a random vector 
$X$ with the associated density $f$ and $\mathcal{F}_q[f](\xi)$ is continuous at $\xi = 0$. Then $X_N$ is weakly $q$-convergent to $X$.
\end{thm}

{\it Proof.} Now assume that $f_N$ tends to $f$ in the sense of $q$-convergence. It follows from Lemma \ref{tight} that 
the corresponding sequence of induced probability measures $\mu_N = P(X^{-1})$ is tight. 
It is well known \cite{Billingsley, Durrett} that if the sequence of associated
probability measures of $X_N$ is
 tight then $X_N \Rightarrow X$ and the density of $X$ coincides with $f$. Lemma \ref{weak} yields 
$X_N \stackrel{q}{\Rightarrow} X$.

\section{MULTIVARIATE $q$-CENTRAL LIMIT THEOREMS}

\subsubsection{Direct multivariate $q$-CLT}

A sequence of random vectors $X_N, N=1,2,...,$ is said to be {\it $(q^d_{-1}-q)$-independent} if, for every N=2,3,...,
\begin{equation}
\label{q-independence}
\mathcal{F}_{q^d_{-1}}[X^c_1+...+X^c_N](\xi)=\mathcal{F}_q[X^c_1](\xi) \otimes_q ... \otimes_q \mathcal{F}_q[X^c_N](\xi),
\end{equation} 
where $q^d_{-1}=z^{d,-1}(q)$ and $X_k^c = X_k - \mu_q(X_k), \, k=1,2,...$.

The relation (\ref{q-independence}) can be rewritten as follows. Denote by $f_N$  
the density of $X^c_1+...+X^c_N$. Then  $(q^d_{-1}-q)$-correlation means 
\begin{equation}
\label{q-independence2}
\int_{R^d}e_{q^d_{-1}}^{i x \xi} \otimes_{q^d_{-1}} f_N(x) dx = \mathcal{F}_q[f_{X^c_1}](\xi) \otimes_q ... \otimes_q \mathcal{F}_q[f_{X^c_N}](\xi), \,
N=2,3,....
\end{equation} 
For $q=1$ the condition (\ref{q-independence2}) turns into the usual independence of identically distributed random vectors $X_1,...,X_N$.
If $q \neq 1$, then $(q^d_{-1}-q)$-independence describes a special type of global correlation.

\begin{thm}
Assume $1/2 < q \leq 1+ 2/d$.  
Let 
$X_1, ..., X_N,...$ be a sequence of $(q^d_{-1}-q)$-independent and identically 
distributed $d$-dimensional random vectors with a finite $q$-mean $\mu_q = (\mu_{q}(X_1),...,\mu_q (X_d))$ and a finite 
$(2q-1)$-covariance matrix $\Sigma_{2q-1}$ with the entries $\sigma_{q,i,j} = \mu_{2q-1}[(X_i-\mu_{q,i})(X_j-\mu_{q,j})].$   
\par
Then $Z_N = \frac{ X_1 + ... + X_N -N \mu_{q} }{ (\nu_{2q-1} N)^{\frac{1}{2(1+d-dq)}} } $ 
is $q^d_{-1}$-convergent as $N \rightarrow \infty$ to a $d$-dimensional $q^d_{-1}$-normal distribution with the 
covariance matrix $\beta \Sigma_{2q-1}^{-1}$ where
\begin{equation}
\label{beta}
\beta = \left( \frac{[2-d(q^d_{-1}-1)] C_{q^d_{-1}}^{2(q^d_{-1}-1)}}{4 q  }  \right)^{\frac{1}{1+d-dq^d_{-1}}}, \, \, q=z^d(q^d_{-1}).
\end{equation}

\end{thm}

{\it Proof.}
Noting that $\mu_q(X_1-\mu_q(X_1))=0$, one can assume that $X_1$ is centered, that is $\mu_q = (0,...,0)$.
Let $f$ be the density associated with $X_1$. 
Using the asymptotic expansion of $q$-exponential, namely 
$e_q^a = 1 + a + {q \over 2}a^2 + o(a^2), \, x \rightarrow 0,$
we obtain  
$$\mathcal{F}_q [f](\xi) =  \\
   \int_{R^d} f(x) \left(1 + i (x, \xi) [f(x)]^{q-1}  
- {q \over 2} (x, \xi)^2 [f(x)]^{2(q-1)} + o((x, \xi)^2 [f(x)]^{2(q-1)}) \right)dx =$$
 \begin{equation}
  1 + i \nu_q (\xi,  \mu_{q})  - (q/2) \nu_{2q-1} (\Sigma_{2q-1} \xi, \xi)  + o(|\xi|^2 ), \, |\xi| \rightarrow 0,
\label{step_20}
  \end{equation}
where $(x,\xi)=x_1 \xi_1 +...+ x_d \xi_d$ and
\[
(\xi,  \mu_{q})=\sum_{j=1}^d\xi_j \mu_{q,j}=0 \, \, \mbox{and} \, \,  (\Sigma_{2q-1} \xi, \xi) = \sum_{j,k=1}^d \sigma_{2q-1, j,k}^d \xi_j \xi_k.
\]
Hence, for $F_q [f](\xi)$, we have the estimate
\[
\mathcal{F}_q [f](\xi) = 1 - (q/2) \nu_{2q-1} (\Sigma_{2q-1} \xi, \xi)  + o(|\xi|^2 ), \, |\xi| \rightarrow 0.
\]
Further, it is readily seen that, for a given random variable $X$ and a real $\rho > 0$, the equality
$\mathcal{F}_q [\rho X](\xi)=\mathcal{F}_q[X](\rho^{1+d-dq} \xi)$ holds.
This yields
$\mathcal{F}_q( (\nu_{2q-1} N)^{-\frac{1}{2(1+d-dq)}} X_1)=\mathcal{F}_q[f]( \frac{\xi}{ \sqrt{N \nu_{2q-1}}  } ).$ Moreover, 
it follows from 
the $(q_{-1}-q)$-independence of $X_1,X_2,...$ and the associativity of the $q$-product that
\begin{equation}
\label{step1000}
\mathcal{F}_{q_{-1}}[Z_N](\xi)= \mathcal{F}_q[f]( \frac{\xi}{ \sqrt{N \nu_{2q-1}}  } ) 
{\otimes_q ... \otimes_q} 
\mathcal{F}_{q}[f]( \frac{\xi}{ \sqrt{N \nu_{2q-1}} } ) \,\,(N\,\mbox{factors}).
\end{equation}
Hence, making use of properties of the $q$-logarithm, from (\ref{step1000}) 
we obtain 
\[
\ln_q \mathcal{F}_{q^d_{-1}}[Z_N](\xi)= N \ln_q \mathcal{F}_q[f]( \frac{\xi}{ 
\sqrt{N \nu_{2q-1}}\sigma_{2q-1}  } ) =
N \ln_q ( 1- \frac{q}{2} \frac{  (\Sigma_{2q-1} \xi, \xi)  }{N} + o(\frac{|\xi|^2}{N})) =
\]
\begin{equation}
\label{step101}
-\frac{q}{2} (\Sigma_{2q-1} \xi, \xi) + o(1), \, N \rightarrow \infty \,,
\end{equation}
locally uniformly by $\xi$. 
\par
Consequently, locally uniformly by $\xi,$
\begin{equation}
\label{step_50}
\lim_{N \rightarrow \infty} \mathcal{F}_{q^d_{-1}}(Z_N) = e_q^{-(q/2) (\Sigma_{2q-1} \xi, \xi) } \in 
\mathcal{G}_q.
\end{equation}
Thus, $Z_N$ is $q$-convergent to the random vector $Z$, whose $q^d_{-1}$-Fourier transform is 
$e_q^{-(q/2) (\Sigma_{2q-1} \xi, \xi) }$. 
 
\par
In accordance with Corollary \ref{cormain}, $Z$ is $q^d_{-1}$-normal, and hence, its 
density is a $q^d_{-1}$-Gaussian with some covariance matrix $\Sigma$ and $\beta$, i.e.  
$$
\mathcal{F}_{q^d_{-1}}[Z](\xi) =  \mathcal{F}_{q^d_{-1}}[G_{q^d_{-1}}(\Sigma; x)]= e_q^{-(q/2) (\Sigma_{2q-1} \xi, \xi) }.
$$ 
Let us now find $\Sigma$ and $\beta$. 
It follows from Corollary \ref{cormain} (see (\ref{image})) that 
$\Sigma = \Sigma_{2q-1}^{-1}$ and (see (\ref{alpha}))
\[
\frac{q}{2} = \frac{[2-d(q^d_{-1}-1)]\,(C_{q^d_{-1}}(\beta))^{2(q^d_{-1}-1)}  }{8 \beta}. 
\]
Solving this equation with respect to $\beta$ we obtain (\ref{beta}).

\begin{thm}
\label{kclt}
Assume a sequence $q^d_{k}, k=0, \pm1,...$ is given as (\ref{qdn}).   Let 
$X_N, \, N=1,2,...,$ be a sequence of $(q^d_{k-1}-q^d_k)$-independent for some $k \in \mathcal{Z}$ and identically 
distributed $d$-dimensional random vectors with a finite $q^d_k$-mean 
and a finite second $(2q^d_k-1)$-variance $\Sigma_{2q^d_k-1}$.

Then $Z_N = {D_{N}(q^d_k)} (X_1 + ... + X_N -N \mu_{q^d_k})$ with 
$D_N(q^d_k)=(N \nu_{2q^d_k-1})^{-\frac{1}{2(2-q^d_k)}}$ is $q^d_{k-1}$-convergent to a $q^d_{k-1}$-normal 
distribution with the covariance matrix $\beta_k \Sigma_{2q^d_k-1}^{-1}$ where
\begin{equation}
\label{betaka}
\beta_k = \left( \frac{[2-d(q^d_{k-1}-1)] C_{q^d_{k-1}}^{2(q^d_{k-1}-1)}}{4 q^d_k  }  \right)^{\frac{1}{1+d-dq^d_{k-1}}}.
\end{equation}

\end{thm}

\begin{remark}
The formulation of Theorem \ref{kclt} changes if we change the definition of $q$-independence.
The current formulation, where the definition of $(q_{k-1}-q_{k})$-independence is given by 
(\ref{q-independence}), corresponds to the first raw of Table \ref{table}. 
The second and third raws reflect the cases when $q$-independence is replaced by 
$(q_{k}-q_{k})$- and $(q_{k-1}-q_{k-1})$-independence respectively. Consequently,
the outcome, that is, the type of convergence and the parameter $\beta_k$ of the
attractor are also changed.  
In all cases  the attractor is $q^d_{k-1}$-Gaussian (see (\ref{qgau})) with 
different values of parameter $\beta=\beta_k$. 
Note that all three cases generalise the classic multivariate
central limit theorem, recovering it if $q_k=1.$
We note also that, in the case $d=1$, the outcome of the 
third raw of Table \ref{table} recovers the result presented in \cite{UmarovTsallisSteinberg}.
\end{remark}

\begin{table}[h]
{\small
\centering
\caption{Interrelation between the type of correlation, convergence and the parameter of the attractor in the
multivariate $q$-CLT.
}
\vspace{.5cm}

\begin{tabular}{|l|l|l|l|} \hline
 
&
{\bf Type of correlation} & 
{\bf Convergence} & 
{\bf Gaussian parameter } \\ \hline

{ 1 } 
& 
       { $F_{q^d_{k-1}}[X+Y]=F_{q^d_k}[X] \otimes_{q^d_k} F_{q^d_k}[Y]$ } & 
       { $q^d_{k-1}-convergent$ } & 
       { $\beta_k = \left( \frac{[2-d(q^d_{k-1}-1)] C_{q^d_{k-1}}^{2(q^d_{k-1}-1)}}{4 q^d_k  }  
\right)^{\frac{1}{1+d-dq^d_{k-1}}}$ } \\ \hline

{ 2 } 
&

         {$F_{q^d_{k}}[X+Y]=F_{q^d_k}[X] \otimes_{q^d_k} F_{q^d_k}[Y]$ } & 
         { $q^d_{k}-convergent$ } & 
         {$\beta_k = \left( \frac{[2-d(q^d_{k-1}-1)] C_{q^d_{k-1}}^{2(q^d_{k-1}-1)}}{4 q^d_k  }  
\right)^{\frac{1}{1+d-dq^d_{k-1}}}$  } \\ \hline

{ 3 }  
&

         {$F_{q^d_{k-1}}[X+Y]=F_{q^d_{k-1}}[X] \otimes_{q^d_k} F_{q^d_{k-1}}[Y]$ } & 
         { $q^d_{k-1}-convergent$ } & 
         {$\beta_k = \left( \frac{[2-d(q^d_{k-1}-1)] C_{q^d_{k-1}}^{2(q^d_{k-1}-1)}}{4 q^d_{k-1}  }  
\right)^{\frac{1}{1+d-dq^d_{k-1}}}$  } \\ \hline

\end{tabular}
\label{table}
}
\end{table}

\subsubsection{Sequential multivariate $q$-CLT}
In this section we consider random vectors $X_N$ with symmetric densities. By 
definition a sequence of random vectors $X_N, \, N=1,2, ...,$ is said to be $(Q_k)$-independent  if 
\begin{equation}
\label{Q-independence}
\mathcal{F}_{Q_k}[X_1+...+X_N](\xi)=\mathcal{F}_{Q_k}[X_1](\xi) \otimes_{q_{k+d-1}} ... \otimes_{q_{k+d-1}} \mathcal{F}_{Q_k}[X_N](\xi),
\forall N=2,3,...,
\end{equation} 
where $\mathcal{F}_{Q_k}$ is $d$-dimesional $Q_k$-sequential Fourier transform and  $Q_k=\{q_{k},q_{k+1},...,q_{k+d-1}\}$.

The relation (\ref{Q-independence}) can be rewritten as 
\begin{equation}
\label{Q-independence2}
\mathcal{F}_{Q_k} [f_N] (\xi) = \mathcal{F}_{Q_k}[f_{X_1}](\xi) \otimes_{q_{k+d}} ... \otimes_{q_{k+d}} \mathcal{F}_{Q_k}[f_{X_N}](\xi), \,
N=2,3,....
\end{equation} 
through the joint density function $f_N$ of $X_1+...+X_N$ and density functions $f_{X_j}$ of $X_j$.  
Let $f(x)$ be a density. Identify consequently the functions $g_0(x_1,...,x_d)$, $g_1(x_1,...,x_{d-1}, \xi_d)$, ... ,$g_d(\xi_1,...,\xi_d)$:
\[
g_0(x)=f(x),
\]
\vspace{.5cm}
\[
g_1(x_1,...,x_{d-1}, \xi_d)= 
\int_{R_{\xi_d}^1} f(x) \, e_{q_{k+d-1}}^{i x_d \xi_d [g_0]^{q_{k+d-1}-1}} dx_d, 
\]
\vspace{.5cm}

$g_j(x_1,...,x_{d-j},\xi_{d-j+1},...,\xi_d) = $
\vspace{.1cm}

\begin{equation}
\label{gj}
\int_{R^{j}} f(x) e_{q_{k+d-1}}^{i x_d \xi_d [g_0]^{q_{k+d-1}-1}} 
\cdot ... \cdot e_{q_{k+d-j}}^{i x_{d-j+1} \xi_{d-j+1} [g_{j-1}]^{q_{k+d-j}-1}} d x_{d-j+1}...dx_d, \, j=2,...,d.
\end{equation}
\vspace{.5cm}

It is not hard to see that
\vspace{.2cm}

\hspace{1cm} $
\mathcal{F}_{Q_k} [f] (\xi) = g_d (\xi_1,...,\xi_d)=
$
\begin{equation}
\label{gd}
\int_{R^{d}} f(x) e_{q_{k+d-1}}^{i x_d \xi_d [g_0]^{q_{k+d-1}-1}} 
\cdot ... \cdot e_{q_{k}}^{i x_{1} \xi_{1} [g_{d-1}]^{q_{k}-1}} d x_{1}...dx_d.
\end{equation}

Denote by $F^m_{d-j}(x_1,...,x_{d-j})$ the marginal density of $f(x)$, i.e.
$$
F^m_{d-j}(x_1,...,x_{d-j}) = \int_{R^j} f(x) d x_{d-j+1}...dx_d.
$$
It folows from (\ref{gj}) that

\hspace{2cm}$g_j(x_1,...,x_{d-j},\xi_{d-j+1},...,\xi_d) =$
\begin{equation}
\label{asympt}
 F^m_{d-j}(x_1,...,x_{d-j}) + o(|\xi_{d-j+1}| + ...+ |\xi_d|), \, \, \, |\xi_{d-j+1}|+...+|\xi_d| \rightarrow 0.
\end{equation}
Further denote 
\[
\sigma_j^2=\int_{R^d} x_{d-j}^2 f(x) \prod_{l=1}^j [ F^m_{d-l}(x_1,...,x_{d-l}) ]^{q_{k+d-j}-1} dx, \, j=1,...,d.
\]
\begin{thm}
\label{seqclt}
Assume $1/2 < q \leq 1+2/d$.
Let $X_1, ..., X_N,...$ be a sequence of $Q_{k}$-independent and identically 
distributed $d$-dimensional random vectors with a symmetric density $f(x)$ and a finite $\sigma_j^2, \, j=1,...,d$.
\par
Then $Z_N = (a_N, Y_N )$, where $Y_N=X_1 + ... + X_N$ and 
$a_N=(a_1,...,a_d),\, a_j= (\sqrt{q_{k+j-1}N}\sigma_j)^{-\frac{1}{2-q_{k+j-1}} }$,
is $q_{k}$-convergent as $N \rightarrow \infty$ to a $d$-dimensional $q_{k-1}$-normal distribution with a
diagonal covariance matrix. 
\end{thm}

{\it Proof.}
It follows from (\ref{gd}) and (\ref{asympt}) that 
\[
\mathcal{F}_{Q_k} [f] (\xi) = 1 - \frac{1}{2}\sum_{j=1}^d q_{k+j-1} \sigma_{j}^2 \xi_j^2 + o(|\xi|^2).
\]
Further, it is readily seen that, for a given random variable $X$ and a real $d$-dimensional vector $\rho > 0$, with positive
coordinates $\rho_j, \, j=1,...,d$,  there holds
$\mathcal{F}_{Q_k} [(\rho, X)](\xi)=\mathcal{F}_{Q_k}[X]( \rho_1^{2-q_{k+d-1}} \xi_1,..., \rho_d^{2-q_{k}} \xi_d)$.
It follows from this relation that
$\mathcal{F}_{Q_k}[ (a_N , X_1 )] = \mathcal{F}_{Q_k}[f]( \frac{ \xi}{ \sqrt{N }  } ).$ Moreover, 
it follows from 
the $Q_k$-independence of $X_1,X_2,...$ and the associativity of the $q$-product that
\begin{equation}
\label{step100}
\mathcal{F}_{Q_k}[Z_N](\xi)= \mathcal{F}_{Q_k}[f]( \frac{\xi}{ \sqrt{N}} ) 
\otimes_{q_{k+d}} ... \otimes_{q_{k+d}} 
\mathcal{F}_{Q_{k}}[f]( \frac{\xi}{ \sqrt{N } }) \,\,(N\,\mbox{factors}).
\end{equation}
Hence, making use of properties of the $q$-logarithm, from (\ref{step100}) 
we obtain 
\[
\ln_{q_{k+d}} \mathcal{F}_{Q_k}[Z_N](\xi)= N \ln_{q_{k+d}} \mathcal{F}_{Q_k}[f]( \frac{\xi}{ 
\sqrt{N } } ) =
N \ln_{q_{k+d}} ( 1- \sum_{j=1}^d \frac{ q_{k+j-1} \sigma^2_j \xi_j^2  }{2N} + o(\frac{|\xi|^2}{N})) =
\]
\begin{equation}
\label{step101}
- \sum_{j=1}^d \frac{ q_{k+j-1} \sigma^2_j }{2} \xi_j^2 + o(1), \, N \rightarrow \infty \,,
\end{equation}
locally uniformly by $\xi$. 
\par
Consequently, locally uniformly by $\xi,$
\begin{equation}
\label{step_50}
\lim_{N \rightarrow \infty} \mathcal{F}_{Q_k}(Z_N) = e_{q_{k+d}}^{- \sum_{j=1}^d \frac{ q_{k+j-1} \sigma^2_j }{2} \xi_j^2 }.
\end{equation}
Thus, $Z_N$ is $Q_k$-convergent to the random vector $Z$, whose $Q_k$-Fourier transform is 
$ e_{q_{k+d}}^{- \sum_{j=1}^d \frac{ q_{k+j-1} \sigma^2_j }{2} \xi_j^2 }$. 
 
\par
Now we show that for $Z$ there exists $q_{k}$-Gaussian 
with some covariance matrix $\Sigma$ such that   
$$
\mathcal{F}_{Q-k}[Z](\xi) =  \mathcal{F}_{Q_k}[G_{q_{k}}(\Sigma; x)]= e_{q_{k+d}}^{- \sum_{j=1}^d \frac{ q_{k+j-1} \sigma^2_j }{2} \xi_j^2 }.
$$ 
In accordance with Corollary \ref{Qkcor} the covariance matrix $\Sigma$ is diagonal with the diagonal entries $\beta_j, \, j=1,...,d,$,
uniquely determined from the system of equations
\[
\frac{ \gamma_{d-j+1} C_{d-j}^{q_{k+d-1}-1} } {4 \alpha_{d-j} \beta_j}  = \frac{ q_{k+j-1} \sigma^2_j }{2}, \, \, j=1,...,d,
\]
where $\gamma_j$ and $\alpha_j, \, j=1,...,d,$ are given by (\ref{alfagama}). The proof of the theorem is complete.

\begin{remark}
The sequential version of the q-CLT formulated above is valid for the general case of non-diagonal covariance matrices.
\end{remark}

\section{Final remarks}

{\it The $q$-triplet and scaling rate.} \\
In nonextensive statistical mechanics, the $q$-triplet 
$(q_{sen},q_{rel},q_{stat})$ has been introduced \cite{Tsallistriplet}, where $sen$, $rel$ and $stat$ stand for {\it sensitivity to the
initial conditions}, {\it relaxation}, and {\it stationary state}, respectively. This triplet describes important features of some complex systems, such as the fluctuating magnetic field of the plasma within the solar wind, as observed in the data sent by the Voyager 1  \cite{Burlagatriplet}. In paper \cite{UmarovTsallisSteinberg} we considered a different triplet, namely $(q_{k-1},q_k,q_{k+1}),$
consisting of three succesive members of the sequence $\{q_k, \, k =0,\pm1,\pm2,...\},$ corresponding respectively 
to the index of the $q$-Gaussian {\it attractor} ($q_{k-1}$), the type of {\it correlation}, or {\it $q$-independence}, ($q_k$), and the {\it scaling rate} ($q_{k+1}$). It appears in fact that there is not just one or other triplet which is relevant, but the entire infinite countable family $\{q_k\}$, each member corresponding to some physical quantity. The one-to-one association of each member of the family to a concrete physical quantity is still an open question; some knowledge is however available (see footnote in \cite{TsallisGellmannSato}). The situation seems to follow the same pattern for multivariate processes as well. The direct version of the multivariate $q$-central
limit theorem (Theorem 4.2) provides information about the attractor, the type of global correlation, and the scaling rate.
More precisely, if the correlation is identified by the index $q_k^d$, then the attractor is a $q_{k-1}^d$-Gaussian, while
the scaling rate is $1/(1+d-dq_{k-1}^d)$ (see, the power of $N$ in the scaling factor in Theorem 4.2, i.e., Eq. (\ref{betaka}). It
follows from Lemma 2.5 that $1/(1+d-dq_{k-1}^d)=q_{k+1}^d.$ Thus, in the $d$-dimensional case, we have the triplet
$(q^d_{k-1},q^d_k,q^d_{k+1})$. If $d=1$ we recover the result of \cite{UmarovTsallisSteinberg}. 
We note that $q_k^d$ is a subsequence of $q_k \equiv q_k^1$ , choosing the members numbered $nd, n=0,\pm1,...$, 
and omitting $(d-1)$ members for each value of $n$.\\

\noindent
{\it Scale invariance.} \\
One important concept in statistical mechanics is the concept of exact or
asymptotic {\it scale-invariance} (or {\it scale-freedom}). In the present context, we may define this as the case 
where the joint distribution (depending on $Nd$ real variables) of a
system made of $N$ particles has (either exactly, or asymptotically for large $N$), as marginal distribution, 
the joint distribution (depending on $(N-1)d$ real variables) of a system made of $(N-1)$ 
particles (see \cite{TsallisGellmannSato,MendesTsallis2001}). 
We address here the study of a similar problem, but related to the dimension $d$. More precisely, we focus on 
the scale invariance problem for a rescaled process of $X_1+...+X_N$ where
$X_1, ... , X_N$ are a sequence of identically distributed $Q_k$-independent (or $(q_{k-1}-q_k)$-independent) 
random vectors, which satisfy the conditions of Theorem \ref{seqclt} (Theorem \ref{kclt} respectively). 
Assume $q_k \geq 1$ and $H_N(x_1,x_2,...,x_d)$ is the density of 
$Z_N = (a_N, Y_N )$, $a_N=(a_1,...,a_d),\, a_j= (\sqrt{q_{k+j-1}N}\sigma_j)^{-\frac{1}{2-q_{k+j-1}} }$,
(or
$Z_N = {D_{N}(q^d_k)} (X_1 + ... + X_N -N \mu_{q^d_k})$ with 
$D_N(q^d_k)=(N \nu_{2q^d_k-1})^{-\frac{1}{2(2-q^d_k)}}$). Then, in accordance with Theorem \ref{seqclt}
(Theorem \ref{kclt}),
$H_N(x_1,...,x_d) \rightarrow G_{d,q_{k-1}}(x_1,...,x_d)$ in the sense of weak convergence. 
It follows  from Lemma \ref{l2} and Corollary \ref{Qkcor} that 
\[
lim_{N \rightarrow \infty} \int_{-\infty}^{\infty} H_N(x_1,x_2,...,x_d)dx_d 
 = \int_{-\infty}^{\infty} G_{d,q_{k-1}}(x_1,x_2,...,x_d)dx_d 
 = G_{d-1,q_{k}} (x_1,...,x_{d-1}),
\]
where $G_{d-1,q_{k}} (x_1,...,x_{d-1})$ is a $(d-1)$-dimensional $q_{k}$-Gaussian.  Hence, for large $N$ the probability density
function of the system $Z_N$ and its $(d-1)$-dimensional marginal density have the same form of $q$-generalized Gaussians with 
the indices $q_{k-1}$ and $q_{k}$ respectively. In other words $H_N(x_1,x_2,...,x_d)$ is, in this specific sense, asymptotically
scale invariant. This property enlightens the applicability of the present concepts and theorems to many 
natural and artificial systems, where $q$-Gaussians are observed.\\

We acknowledge enlightening remarks by C. Vignat at the early stages of the present work. One of us (C.T.) has benefited from interesting discussions with S.M.D. Queiros. Partial financial support 
from Pronex/MCT, CNPq and Faperj (Brazilian Agencies) is acknowledged as well.

\end{document}